\documentclass[showpacs,amsmath,amssymb]{emulateapj}
\usepackage{graphics}
\usepackage{epsfig}
\usepackage{natbib}
\usepackage{amsmath}
\begin{document}
\title{A Conditional Luminosity Function Model of the Cosmic Far-Infrared Background Anisotropy Power Spectrum}
\author{Francesco De Bernardis, Asantha Cooray}
\affil{Department of Physics and Astronomy, University of California, Irvine, CA 92697}

\begin{abstract}
The cosmic far-infrared background (CFIRB) is expected to be generated by faint, dusty star-forming galaxies during the peak epoch of galaxy formation. The anisotropy power spectrum of the CFIRB
captures the spatial distribution of these galaxies in dark matter halos and the spatial distribution of dark matter halos in the large-scale structure.
Existing halo models of CFIRB anisotropy power spectrum are either incomplete or lead to halo model parameters that are inconsistent with the galaxy distribution
selected at other wavelengths. Here we present a conditional luminosity function approach to describe the far-IR bright galaxies.
We model the 250 $\mu$m luminosity function and its evolution with redshift
and model-fit the CFIRB power spectrum at 250 $\mu$m measured by the Herschel Space Observatory.
We introduce a redshift dependent duty-cycle parameter so that we are able to
estimate the typical duration of the dusty star formation process in the dark matter halos as a function of redshifts.
We find the duty cycle of galaxies contributing to the far-IR background is 0.3 to 0.5 with
a dusty star-formation phase lasting for $\sim0.3-1.6$ Gyrs. This result confirms the general expectation that the far-IR
background is dominated by star-forming galaxies in an extended phases, not bright starbursts that are driven by galaxy mergers
and last $\sim10-100$ Myrs. The halo occupation number for satellite galaxies has a power-law slope that is close to unity over $0 < z < 4$.
We find that the minimum halo mass for dusty, star-forming galaxies with
$L_{250} > 10^{10}$ L$_{\sun}$ is $2 \times 10^{11}$ M$_{\odot}$ and $3 \times 10^{10}$ M$_{\odot}$ at $z=1$ and 2, respectively.
Integrating over the galaxy population with $L_{250} > 10^{9}$ L$_{\sun}$, we find that the cosmic density of dust residing in the
dusty, star-forming galaxies responsible for the background anisotropies $\Omega_{\rm dust} \sim 3\times10^{-6}$ to $2 \times10^{-5}$, relative to the
critical density of the Universe.
\end{abstract}

\section{Introduction}

The total intensity of the cosmic far-infrared background (CFIRB) is now established with absolute photometry \citep{Puget1996,Fixen1998,Dwek1998}.
This background originates from the UV and optical emission of young stars, absorbed by the dust in galaxies and then re-emitted in the infrared (IR) wavelengths.
Deep surveys with instruments aboard the {\it Herschel} Space Observatory have started to resolve this background intensity
between 100 and 500 $\mu$m to discrete galaxies based on resolved counts \citep{hermes,Clements2010,Berta:2011xi}.
Unfortunately even the deepest images of the far-IR sky using PACS and SPIRE are limited by source confusion.
For example, at 250, 350, and 500 $\mu$m, only  15, 10 and 6\% of the total background intensity is resolved to individual galaxies, respectively. Instead of individual detections,
the fainter galaxies responsible for the bulk of the CFIRB intensity is studied with statistics such as $P(D)$, the probability of deflections \citep{Glenn2010},
and $P(k)$, the angular power spectrum of CFIRB anisotropies resulting from the correlated confusion noise \citep{Haiman:1999hh,Knox2001,Scott:1998ei,Negrello:2007kv,Amblard:2007zq}.

While attempts were made to detect the power spectrum of CFIRB with {\it Spitzer}-MIPS at 160 $\mu$m \citep{iras},
and a limited low signal-to-noise ratio detection with BLAST \citep{Devlin} in \citet{Viero2009}, the first clear detection of
the CFIRB anisotropy power spectrum from 30 arcseconds to 30 arcminute angular scales came from {\it Herschel}-SPIRE at 250, 350 and 500 $\mu$m \citep{Amblard:2011gc}.
This was soon followed by Planck measurements of the CFIRB power spectrum from 5 arcminute to degree angular scales \citep{Ade:2011ap}. At the longer mm-wavelengths, clustering of dusty galaxies can also be studied as part of the CMB secondary anisotropy studies, where a combination of
signals contribute to the total power spectrum \citep{Addison:2011se,Archidiacono:2011eh}.
While the arcminute-scale ground-based CMB experiments and Planck can study the large-angular correlations in the
CFIRB at linear scales, the angular resolution of {\it Herschel}-SPIRE \citep{spire} is such that the measurements probe the non-linear scales
and capture important information on how the dusty star-forming galaxies are distributed in the dark matter halos.

First predictions on the CFIRB anisotropy power spectrum concentrated on the linear power spectrum scaled by a bias factor \citep{Scott:1998ei,Haiman:1999hh,Knox2001}.
Since those early studies a popular approach to describe the large-scale structure galaxy distribution is to connect galaxies to the
underlying dark matter halo distribution  (see review in \citealt{Cooray:2002dia}). This halo modeling allows a way to describe the galaxy clustering power
spectrum and correlation function through the halo occupation number describing the number of galaxies in a given dark matter halo as a function of the halo mass. Recent
improvements in the halo model involve an occupation number description that takes into
account for the luminosity dependence of the satellites
through what are now called conditional luminosity functions (CLFs; \citealt{Yang2004,Cooray:2005yt,Cooray:2006di}).

While an attempt was made to incorporate CLFs to describe the CFIRB power spectrum (\citealt{Amblard:2007zq}; see recent works in \citealt{Shang:2011mh,Wang:2010ik,Xia2012}),
this was based on a phenomenological model for the number counts and luminosity functions (LFs) of far-IR sources \citep{Lagache:2002xq}. The number counts and LF
measurements from the {\it Herschel} Space Observatory now allow us to both improve the model and extract parameters of the underlying CLF description.

Separately, modeling of recent measurements of the CFIRB anisotropy power spectrum with {\it Herschel} and Planck, and the dusty galaxy signal in CMB secondary anisotropy data,
is somewhat controversial. The best-fit parameters of the original study \citep{Amblard:2011gc} either had a power-law slope for satellites that was steeper than 1.3 or had a relation between the
satellite mass scale $M_{\rm sat}$ and the minimum halo mass to host a galaxy $M_{\rm min}$ such that $M_{\rm sat} \sim (3-4)M_{\rm min}$.
The galaxy clustering measurements in the optical band show that the power-law slope is slightly less than 1 \citep{Zehavi2004,Abazajian:2004tn},  while  $M_{\rm sat} \sim (15-20)M_{\rm min}$ \citep{Gao:2004au, Kravtsov:2003sg, Zheng:2004id, Hansen:2007fy, Shang:2011mh}.
The issue is not limited to the {\it Herschel} power spectrum since similar conclusions can also be reached with fits to the Planck CFIRB power spectrum.
Prior to {\it Herschel}, the low signal-to-noise
CFIRB power spectrum reported by BLAST \citep{Viero2009} required a halo profile that extends out to $\sim 4 r_{\rm vir}$ to fit the data, leading to an overestimate of
the mean density of dark matter in the universe relative to the value in $\Omega_m$ that normalizes the dark matter halo mass function.

While a power-law description of the CFIRB
power spectrum out to $\ell \sim 2000$ is likely adequate for the dusty galaxy power spectrum in ground-based arcminute-scale CMB anisotropy data \citep{Addison:2011se},
a clear departure from the power-law
was detected in the {\it Herschel} measurement out to $\ell> 10^4$, indicating the transition between the 2-halo and 1-halo term of galaxy clustering. A proper description of
the {\it Herschel} CFIRB power spectrum must then move beyond the power-law fit to the data.
We also refer the reader to \citet{Penin2012,Bethermin2012} for more recent modeling of CFIRB,
concentrating on the Planck-measured CFIRB power spectrum, and \citet{Xia2012} for modeling of both Planck and {\it Herschel}.

This paper is organized as follows: in the next Section we outline the {\it Herschel} data used for this analysis. In Section~3 we present a revised CLF model for the CFIRB anisotropy
power spectrum. In Section~4 we present our results and conclude with a summary in Section~5.
Throughout this paper we assume the fiducial cosmology for the $\Lambda$CDM model of WMAP-7 results \citep{Komatsu:2010fb}.

\section{Data used for the analysis}

The CFIRB angular power spectrum used for this analysis  is the same as that of \citet{Amblard:2011gc}, taken from the {\it Herschel} Multi-tiered Extra-galactic survey \citep{hermes} with the Spectral and Photometric Imaging Receiver \citep{spire} onboard the {\it Herschel} Space Observatory \citep{herschel}. While the measurements were reported for three wavelengths, we concentrate on
the 250 $\mu$m angular power spectrum since it has the highest signal-to-noise and the best resolution.

The luminosity function data measured by {\it Herschel} are taken from \citet{Vaccari:2010vu} at low redshifts ($z<0.2$) at
$250\mu m$.  The high-z luminosity function data extending up to $z=4$ at 250 $\mu$m are from
\citet{Eales:2010vw} and  \citet{Lapi:2011ca}. We use data from \citet{Eales:2010vw}  in 2 bins at $0.2<z<0.4$ and $0.4<z<0.8$. These luminosity functions are based out of optical and near-IR photometric or spectroscopic redshifts for 250 $\mu$m-detected galaxies in the GOODS-North field. To extend the 250 $\mu$m
LFs to higher redshifts we make use of the results from \citet{Lapi:2011ca}. These luminosity functions are somewhat uncertain as they are based on the sub-mm photometric redshifts, which for each galaxy could have an error of at least 0.3 in $\Delta z/(1+z)$
\citep{Harris:2012iz}. In any case some of that uncertainty is captured by the errors of the LF.
In future with more exact LFs our model can be further improved.

The low-z luminosity function data are shown in Fig.~\ref{lowz_plots} left, while the CFIRB angular power spectrum data are in Fig.~\ref{highz_plots} left. Fig.~\ref{highz_plots_2} left shows the high redshift luminosity functions data.

\section{Halo-model and Luminosity function formalism}\label{model}

The CLF model we use here to analyze the LF and $P(k)$ data is largely based on the model of \citet{Giavalisco:2000xs} and \citet{Leeetal2009}.
One of the main advantages of this improved description of a halo model is that it clearly connects the luminosity function to clustering of galaxies, allowing to simultaneously constrain
the model parameters using both these observables. The connection between 1-point (LF) and 2-point ($w(\theta)$, $P(k)$) is based on an
 explicit model of the galaxy luminosity-halo mass relation as a function of redshift. When compared to the standard halo model with galaxy statistics
described by an occupation number, such a luminosity based approach is capable of accounting for the fact that the luminous galaxies
are more likely to be in more massive halos.  Moreover, the CLF-based model description of
\citet{Leeetal2009} is  general enough to reproduce a wide range of shapes for the galaxy luminosity-halo mass relation and its scatter.
This is advantageous as the shape of this relation is expected to be different at far-IR wavelengths when compared to the same data at optical wavelengths.

The fundamental ingredients in this revised CLF model are the mass functions for halos and sub-halos and the galaxy luminosity-halo mass relation and its evolution.
The probability density for a halo or a sub-halo of mass $M$ to host a galaxy with luminosity $L$ is modeled as a normal distribution with
\begin{eqnarray}
P(L|M)=\frac{\eta_{\rm DC}}{\sqrt{2\pi}\sigma_L(M)}\exp\left[-\frac{(L-\bar{L}(M))^2}{2\sigma_L(M)^2}\right] \, ,
\label{pml}
\end{eqnarray}
where $\eta_{\rm DC}$ is the {\it duty cycle} factor related to the duration of the star formation in the halos (and is $0\le \eta_{\rm DC}\le1$).
More precisely, the duty cycle represents a measure of the duration of the star formation, $t_{\rm SF}$, relative to the time interval probed by the survey or the
observations, $\Delta t$.  As discussed in \citet{Leeetal2009}, the ratio $t_{\rm SF}/\Delta t$ determines the number of halos that can host a detectable galaxy
and is hence related to the ratio  between the number densities of galaxies and available halos to host such galaxies $n_g/n_h$.
This is precisely the duty cycle $\eta_{\rm DC}$ we have introduced above.

Note that this description of the duty cycle
is different from the ``duty cycle'' reported by \citet{Shang:2011mh} for dusty, starforming galaxies in their halo/CLF modeling of the Planck
CFIRB power spectra. In their work, the duty cycle is derived by comparing the measured shot-noise amplitude, that dominates
anisotropy power spectrum at small angular scales, to a prediction of the expected shot-noise given the number density of halos and the
observed counts. Given that the shot-noise is $\int dS S^2 dn/dS$, where $S$ is the flux density and $dn/dS$ is the number counts, the shot-noise quoted in their
paper is weighted more towards the bright, rare sources. The model comparison by \citet{Shang:2011mh}  suggests a duty cycle that is close to one
suggesting that the CFIRB anisitropies are dominated by normal quiescent galaxies. Here we provide a precise estimate of the duty cycle down to a specific luminosity
and as a function of redshift.

In eq.~\ref{pml}, $\sigma(M)$ is the scatter in the luminosity-mass relation. In this description the scatter can be related to the nature of the starformation.
High values for the  scatter with respect to the mean luminosity $\bar{L}(M)$ imply a star formation dominated by starbursts while low values for scatter, suggesting
a fixed relation between halo mass and luminosity, are typical of quiescent,  steady star formation (see also discussion in \citealt{Leeetal2009}).

The relation between the halo mass and the average luminosity $\bar{L}(M)$ is expected to be an increasing function of the mass with a characteristic mass scale $M_{0l}$ and we can write (see \citet{Leeetal2009})
\begin{eqnarray}\label{lm_relation}
\bar{L}(M)=L_{0}\left(\frac{M}{M_{0l}}\right)^{\alpha_l}\exp\left[-\left(\frac{M}{M_{0l}}\right)^{-\beta_l}\right] \, ,
\end{eqnarray}
and the scatter can be parameterized in a similar way
\begin{eqnarray}\label{scatter}
\sigma(M)=\sigma_{0}\left(\frac{M}{M_{0s}}\right)^{\alpha_s}\exp\left[-\left(\frac{M}{M_{0s}}\right)^{-\beta_s}\right] \, .
\end{eqnarray}

As already discussed by  \citet{Leeetal2009} these parameterizations don't have a specific physical motivation (except for the requirement of being increasing function of mass), but offer the advantage to explore a large range of possible shapes.
We need to consider the total halo mass function, that is the number density  of halos or sub-halos of mass $M$.
The contribution of halos $n_h(M)$  is taken to be the Sheth \& Tormen relation \citep{ShethTormen1999}.
The sub-halos term can be modeled through the number of sub-halos of mass $m$ inside a parent halo of mass $M_p$, $N(m|M_p)$. The total mass function is then
\begin{eqnarray}
n_T(M)=n_h(M)+n_{sh}(M) \, ,
\end{eqnarray}
where $n_{sh}(M)$ is the sub-halo mass function
\begin{eqnarray}
n_{sh}(M)=\int N(M|M_p)n_h(M_p)dM_p \, .
\end{eqnarray}
We parameterize $N(m|M)$ as in \citet{vandenBosch:2004zs}
\begin{eqnarray}
N(m|M)=\frac{\gamma}{\beta\Gamma(1-\alpha_{sh})}\left(\frac{m}{M\beta_{sh}}\right)^{-\alpha_{sh}}\exp\left(-\frac{m}{M\beta_{sh}}\right) \, ,
\end{eqnarray}
where
\begin{eqnarray}\label{norma_sub}
\gamma=\frac{f_{sh}}{\Gamma(1-\alpha_{sh},1/\beta_{sh})-\Gamma(1-\alpha_{sh},10^{-4}/\beta_{sh})} \, .
\end{eqnarray}
Here $\Gamma(x)$ is the incomplete gamma function and $f_{sh}$ is the sub-halo mass fraction.
As shown in \citet{vandenBosch:2004zs}, where the model is calibrated using numerical simulations, both the normalization and the slope of the sub-halo mass function
are not universal and depend on the ratio between the parent halo mass and the non linear mass scale, $M_*$, defined as the mass scale where the rms of the density field $\sigma(M,z)$ is equal to the critical over-density required for the spherical collapse $\delta_c(z)$ . The term $f_{sh}$ in eq.~\ref{norma_sub} is fitted by the relation
\begin{eqnarray}
\log[\langle f_{sh}\rangle]=\left[0.4(\log(M/M_*)+5)\right]^{1/2}+2.74 \, ,
\end{eqnarray}
in numerical simulations and we make use of it in this study.

The best-fit relation for the slope parameters $\alpha_{sh}$ and $\beta_{sh}$ found by \citet{vandenBosch:2004zs} is
\begin{eqnarray}
\alpha_{sh}=0.966-0.028\log(M/M_*) \, ,
\end{eqnarray}
and $\beta_{sh}=0.13$, independent of $M$. With this description the total number of free parameters in the CLF model is $9$ with
$4$ parameters for the luminosity-mass relation in eq.~\ref{lm_relation}, $4$ parameters for the scatter in eq.~\ref{scatter},  and the duty cycle parameter $\eta_{\rm DC}$.

If the same luminosity-mass relation applies to both halos and sub-halos, then the product $P(L|M)n_T(M)dLdM$
gives the number densities of galaxies with luminosity $L$ in halos or sub-halos of mass $M$. The luminosity function is then
\begin{eqnarray}
\phi(L)dL=dL\int dMP(L|M)n_T(M) \, .
\end{eqnarray}

The formalism introduced above also allows us to construct the halo occupation distribution (HOD) in a simple way.
The contribution of central galaxies is simply the integration of $P(L|M)$ over all
luminosities above a certain threshold $L_0$ either fixed by the survey or a priori selected so that
\begin{eqnarray}
\langle N_c(M)\rangle_{L\ge L_{min}}=\int_{L_{min}}P(L|M)dL \, ,
\end{eqnarray}
which, in absence of scatter, reduces to a step function $\Theta(M-M_0)$ as expected. For the satellite galaxies,  the HOD is related to the sub-halos
\begin{eqnarray}
\langle N_{s}(M)\rangle_{L\ge L_{min}}=\int_{L_{min}}dL\int dmN(m|M) P(L|m)\, .
\end{eqnarray}
The total HOD is then
\begin{eqnarray}
\langle N_{\rm tot}(M)\rangle_{L\ge L_{min}}=\langle N_{h}(M)\rangle_{L\ge L_{min}}+\langle N_{sh}(M)\rangle_{L\ge L_{min}} \, .
\end{eqnarray}

The model described so far holds at a given redshift.
The duty cycle parameter $\eta_{\rm DC}$ and the luminosity-mass relation with its scatter are expected to have a redshift evolution.
Here we are attempting to fit LF data at a variety of redshift bins between $ 0 < z < 4$.
 To account for the redshift evolution of the parameters we assume that the parameters that the describe the low redshift ($z< 0.2$)
dusty galaxy population are different from those for the high redshift galaxies.
Moreover, the high-z data extend from $z=0.2$ to $z=4$ and are divided in to $6$ redshift bins. We thus fit a total of
$7$ duty cycle parameters, one for each of the bins and do not attempt to constrain the duty cycle variation with a parameterized approach on the redshift evolution.
For the evolution of the galaxy 250 $\mu$m luminosity-halo mass relation we account for the possible redshift evolution by introducing another parameter, $p_M$
and rewriting the mass scale $M_{0l}$ as
\begin{eqnarray}\label{mol_evolution}
M_{0l}(z)=M_{0l,z<0.2}(1+z)^{p_M} \, ,
\end{eqnarray}
where we allow the evolution to follow the assumed power-law  form.

Once the HOD is defined, it is possible to calculate the one-halo and two-halo terms of the far-IR anisotropy power spectrum. First, we define the power spectrum in terms of
redshift-dependent three-dimensional clustering and will later project them along the line of sight to calculate the angular power spectrum of CFIRB anisotropies.
Here we assume that the central galaxy is at the center of the halo and that the halo radial profile of satellite galaxies within dark matter halos follow that of the
dark matter given by the Navarro, Frenk and White (NFW) profile \citep{NFW}. The one-halo term is then
\begin{eqnarray}\label{1h}
P^{1h}(k)=\frac{1}{n_g^2}\int dM\langle N_T(N_T-1)\rangle u(k,M)^pn_h(M) \, ,
\end{eqnarray}
where $u(k,M)$ is the NFW profile in Fourier space and $n_g$ is the galaxy number density
\begin{eqnarray}
n_g=\int dM\langle N_g(M)\rangle n_h(M) \, .
\end{eqnarray}
The second moment of the HOD that appear in eq.~\ref{1h} can be simplified as
\begin{eqnarray}
\langle N_T(N_T-1)\rangle\simeq\langle N_T\rangle^2-\langle N_h\rangle^2\, ,
\end{eqnarray}
and the power index $p$ for the NFW profile is $p=1$ when  $\langle N_T(N_T-1)\rangle<1$ and $p=2$ otherwise  \citep{Leeetal2009}.
The two-halo term  of galaxy power spectrum is
\begin{eqnarray}
P^{2h}(k)&=&\left[\frac{1}{n_g}\int dM\langle N_T(M)\rangle u(k,M)n_h(M)b(M)\right]^2\\
&&\times P_{\rm lin}(k)\, ,
\end{eqnarray}
where $P_{\rm lin}(k)$ is the linear power spectrum and $b(M)$ is the linear bias factor calculated as in \citet{Cooray:2002dia}. The total galaxy power spectrum is then $P_g(k)=P^{1h}(k)+P^{2h}(k)$.

As the observations are anisotropies on the sky projected along the line of sight,
the observed angular power spectrum can be related to the three-dimensional galaxy power spectrum through a redshift integration along the line of sight \citep{Knox2001}:
\begin{eqnarray}
C_{\ell}^{\nu \nu'}=\int dz\left(\frac{d\chi}{dz}\right)\left(\frac{a}{\chi}\right)^2\bar{j}_{\nu}(z)\bar{j}_{\nu'}(z)P_g(\ell/\chi,z) \, ,
\end{eqnarray}
where $\chi$ is the comoving radial distance, $a$ is the scale factor and $\bar{j}_{\nu}(z)$ is the mean emissivity at the
frequency $\nu$ and redshift $z$ per comoving unit volume that can be obtained from the LFs as
\begin{eqnarray}\label{jnu}
\bar{j}_{\nu}(z)=\int dL\phi(L,z)\frac{L}{4\pi} \, .
\end{eqnarray}
This model does not rely on the assumption of a number counts shape or an evolution. We are able to directly model-fit the mean emissivity as a function of redshift.

\begin{table}[h!]
\begin{center}
\begin{tabular}{lr}
\hline
$\alpha_l$   											&	$0.22\pm0.10$\\
$\beta_l$     											&	$0.70\pm0.05$\\
$\log (M_{0l}/M_{\odot})$						        &	$11.5\pm1.7$\\
$\log (L_{0}/L_{\odot})$							   &	$9.6\pm2.4$\\
$\eta_{\rm DC}$ ($0<z<0.2$)							&	$0.54\pm0.26$\\
\hline
\end{tabular}\caption{Best-fit parameter values and their $1-\sigma$  uncertainties
from the low-redshift luminosity function data of \citet{Vaccari:2010vu}}\label{lowz_table}
\end{center}
\end{table}%

\begin{table}[h!]
\begin{center}
\begin{tabular}{lr}
\hline
$\alpha_l^\prime$   											&	$0.57\pm0.03$\\
$\beta_l^\prime$     											&	$0.19\pm0.02$\\
$\log (L_{0}^\prime/L_{\odot})$							   &	$9.58\pm0.02$\\
$C_{250}$												&	$0.78\pm0.16$\\
$\eta_{\rm DC}$ ($0.2<z<0.4$)														&	$0.43\pm0.07$\\
$\eta_{\rm DC}$ ($0.4<z<0.8$)														&	$0.30\pm0.04$\\
$\eta_{\rm DC}$ ($1.2<z<1.6$)														&	$0.16\pm0.01$\\
$\eta_{\rm DC}$ ($1.6<z<2.0$)														&	$0.19\pm0.01$\\
$\eta_{\rm DC}$ ($2.0<z<2.4$)                                                      &   $0.33\pm0.01$\\
$\eta_{\rm DC}$ ($2.4<z<4.0$)                                                          &   $0.31\pm0.02$\\
$p_M$                                                        &   $-4.32\pm0.09$ \\
\hline
\end{tabular}\caption{Best-fit parameter values and their $1-\sigma$ uncertainties from the combination of angular CFIRB power spectrum
at $250\mu $m and high-z luminosity function data.}\label{highz_table}
\end{center}
\end{table}%

\section{Results and Discussion}

In the revised CLF model outlined above, in principle, we have
  $23$ free parameters: $7$ duty cycle parameters, $8$ parameters for the luminosity-mass relation and its scatter at low redshifts, $7$ parameters for the same relations
at high redshifts plus the parameter $p_M$ for the $(1+z)$ redshift evolution of the mass scale (eq.~\ref{mol_evolution}).

Separately, the CFIRB power spectrum contains the contribution from Galactic cirrus, in addition to the extragalactic anisotropies
traced by the faint, dusty galaxies. In \citet{Amblard:2011gc} the authors accounted for this contamination assuming the same cirrus power-law power-spectrum from measurements of IRAS and
MIPS \citep{iras} at $100\mu m$ and extending it to higher wavelengths using the spectral dependence of \citet{schlegel}. Such a frequency scaling resulted in an overestimated
cirrus correction, as noted by the Planck team \citep{Ade:2011ap} in their analysis of the CFIRB power spectrum compared to the {\it Herschel} power spectrum. This is primarily due to
the fact that the cirrus is likely overestimated in \citet{schlegel} as IRAS 100 $\mu$m also contains the extragalactic background intensity.
To avoid biasing our power spectrum low by an overestimated cirrus correction, we re-fit the raw power spectrum data from \citet{Amblard:2011gc}.
Here we adopt the same power-law cirrus fluctuation power spectrum  used in \citet{Amblard:2011gc}, with $P(k) \propto k^{-n}$ with $n=-2.89\pm0.22$ as measured by \citep{iras}. However, we rescale the amplitude of the cirrus power spectrum with a dimensionless factor $C_{250}$ that we keep as a free parameter and model-fit that as part of the global halo model.
This implies another free parameter in our model, leading to a total of $24$ parameters. Given the large volume of the parameter space, a MCMC analysis (see below) through
the full parameter space is very time-consuming. Moreover, it is unlikely that the information carried by the current data is able to constrain such a large number of free parameters.

\begin{figure*}[htb!]
\begin{tabular}{cc}
\epsfig{file=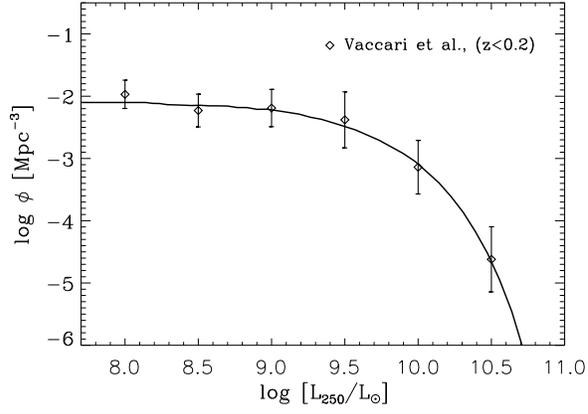,width=0.5\linewidth,clip=}&
\epsfig{file=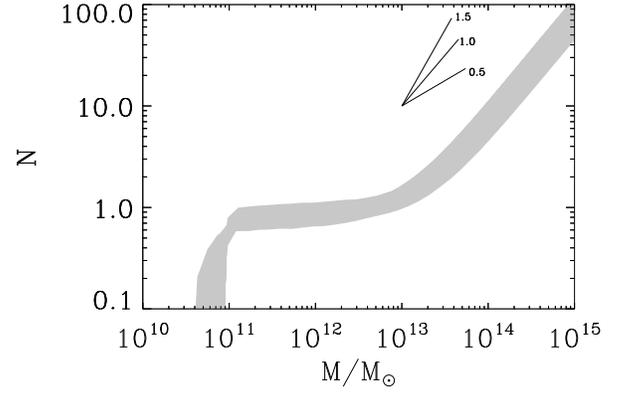,width=0.5\linewidth,clip=}\\
\end{tabular}\caption{{\it Left:} 250 $\mu$m luminosity function data at $z<0.2$ and the  best-fit model. {\it Right:} The halo occupation number at $z<0.2$.
The minimum halo mass to host a galaxy with luminosity $L>5\cdot10^{7} L_{\odot}$ is $\log(M_{\rm min}/M_{\odot})\simeq10.8$ and the power-law slope of the satellite galaxies with halo mass
is  $\sim0.98$. The grey region represents the $68\%$ confidence level and the three lines at the top right of the figure correspond to power-law slopes of 0.5, 1.0 and 1.5.}
\label{lowz_plots}
\end{figure*}

\begin{figure*}[htb!]
\begin{tabular}{cc}
\epsfig{file=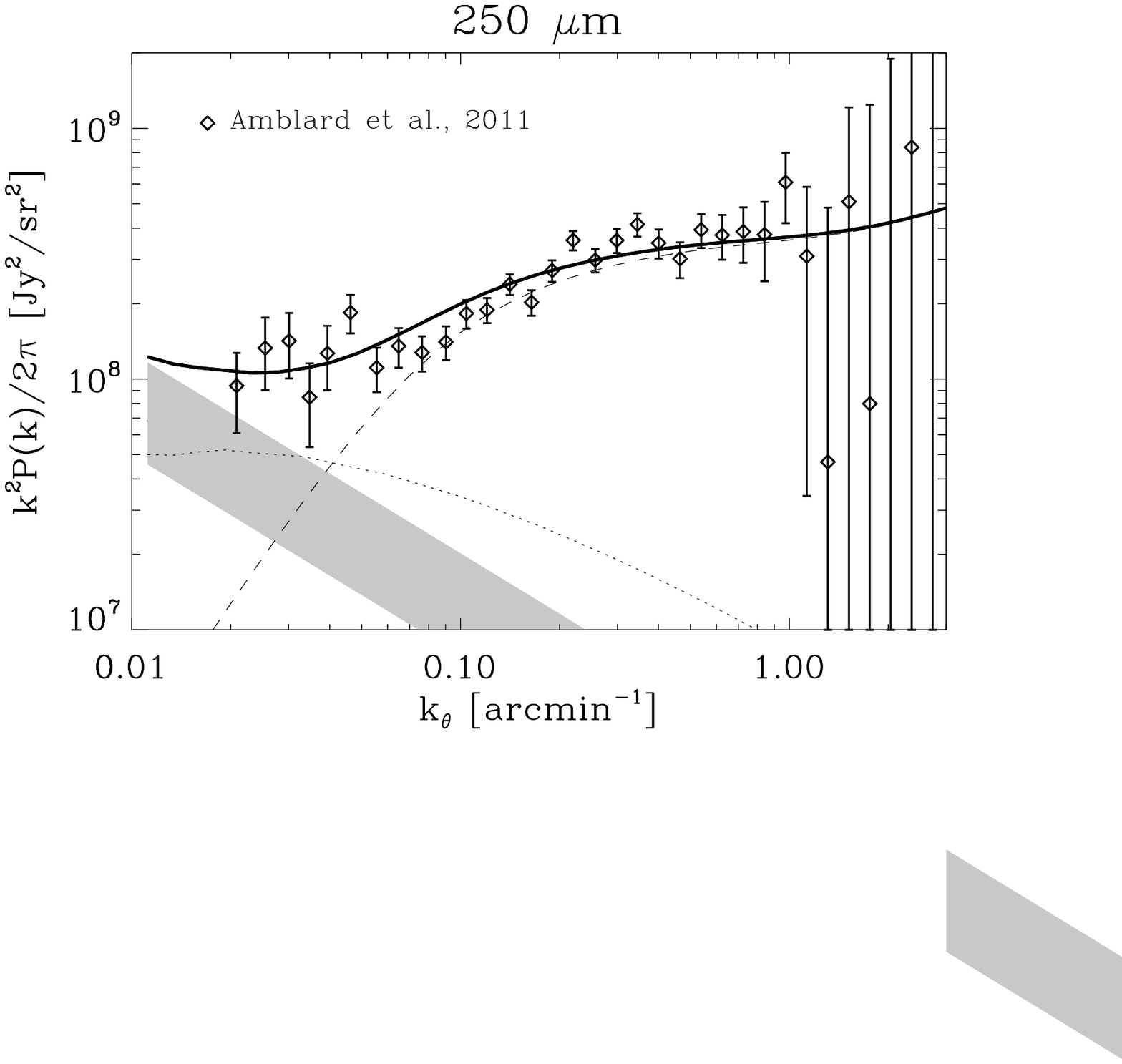,width=0.5\linewidth,clip=}&
\epsfig{file=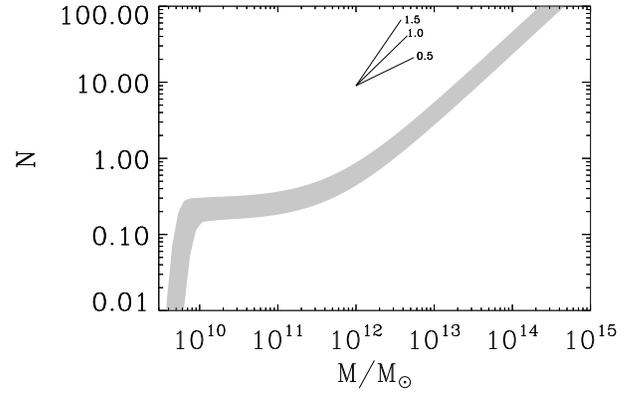,width=0.5\linewidth,clip=}\\
\end{tabular}\caption{{\it Left:} Best-fit model description to the angular power spectrum data at $250 \mu$m. When fitting to the angular power spectrum data
we removed  the shot-noise contribution as determined by \citet{Amblard:2011gc}. Dashed and dotted lines are the $1$-halo and $2$-halo term, respectively.
The shaded region is the $68\%$ confidence level galactic cirrus contribution, as determined by the free parameter $C_{250}$ included in the model-fit.
{\it Right:} The halo occupation number for the high-z galaxies.
The HOD slope at the high mass-end is $\simeq0.96$. The grey region represents $68\%$ confidence level and the three lines correspond
correspond to power-law slopes of 0.5, 1.0 and 1.5.}
\label{highz_plots}
\end{figure*}

\begin{figure*}[htb!]
\begin{tabular}{cc}
\epsfig{file=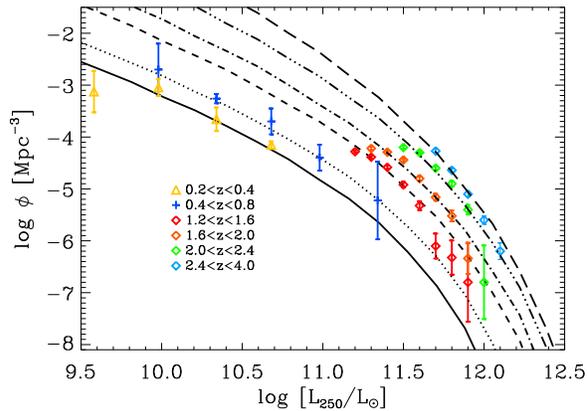,width=0.5\linewidth,clip=}&
\epsfig{file=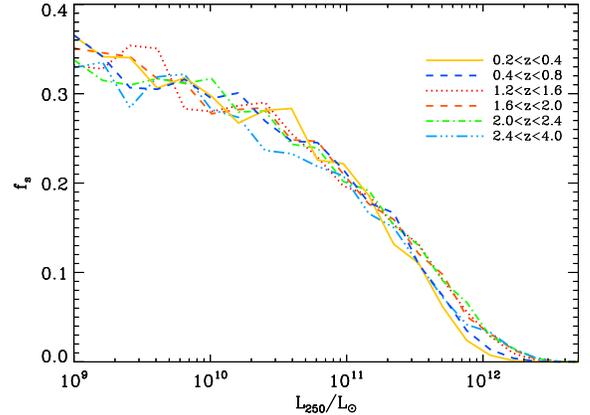,width=0.5\linewidth,clip=}\\
\end{tabular}\caption{{\it Left:} Best-fit model description
to the high-redshift luminosity functions from  \citet{Eales:2010vw} and \citet{Lapi:2011ca}.
{\it Right:} Satellite fraction at each redshift bin.}\label{highz_plots_2}
\end{figure*}

\begin{figure*}[htb!]
\begin{tabular}{cc}
\epsfig{file=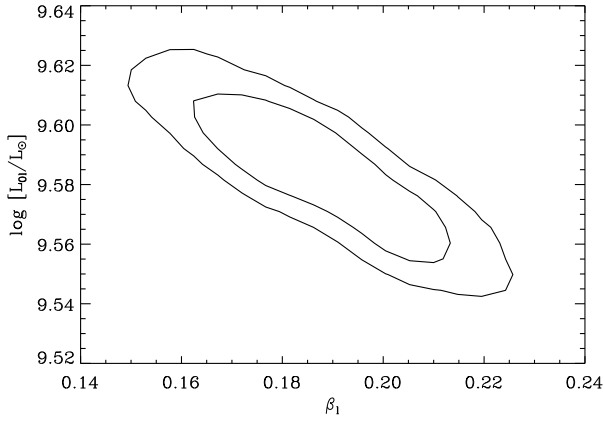,width=0.5\linewidth,clip=}&
\epsfig{file=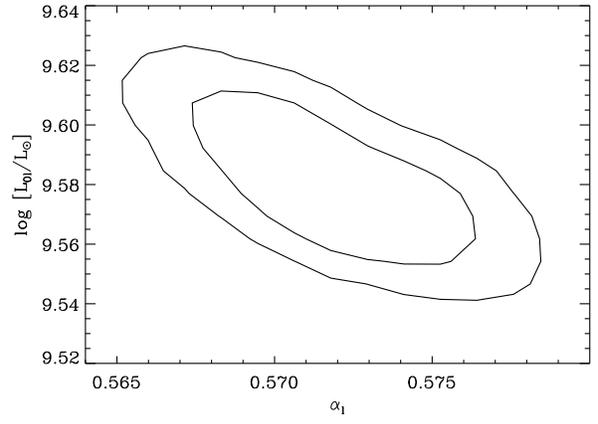,width=0.5\linewidth,clip=}\\
\end{tabular}\caption{$68\%$ and $95\%$ confidence level contours for the parameters of the luminosity-mass relation $L_0^\prime$-$\beta_l^\prime$ ({\it left}) and $L_0^\prime$-$\alpha_l^\prime$ ({\it right}).}\label{contours}
\end{figure*}

\begin{figure}[htb!]
\epsfig{file=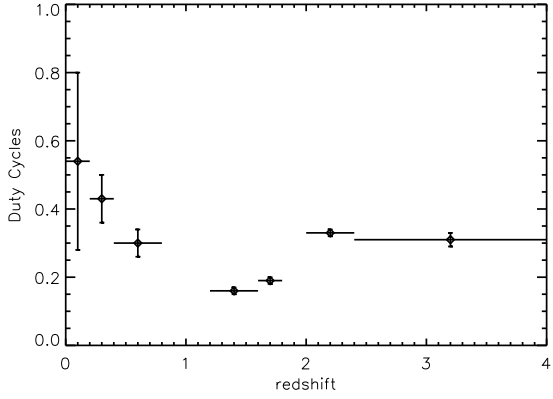,width=1\linewidth,clip=}
\caption	{Duty cycle $\eta_{\rm DC}$ as a function of
redshift. Over the redshift range of $1 < z <4$, $\eta_{\rm DC} \sim 0.2$ to 0.4.
}\label{dc_z}
\end{figure}

\begin{figure}[htb!]
\epsfig{file=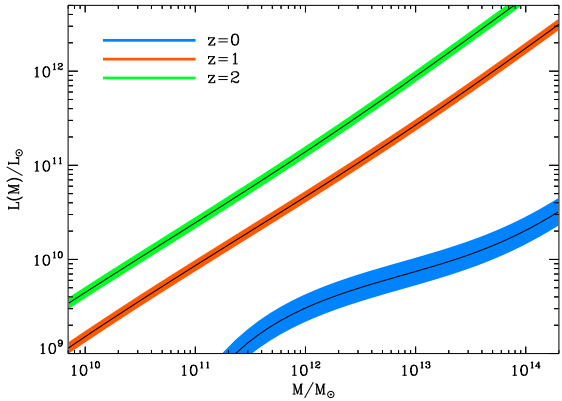,width=1\linewidth,clip=}
\caption{The 250 $\mu$m luminosity-halo mass relation and the $68\%$ confidence level region at $z=0$, 1 and 2.}\label{LM}
\end{figure}

\begin{figure}[htb!]
\epsfig{file=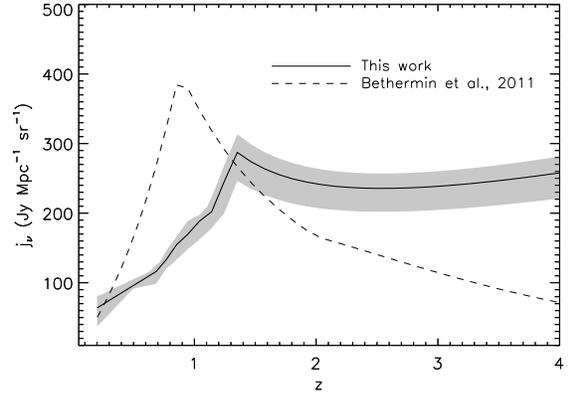,width=1\linewidth,clip=}
\caption{250 $\mu$m emissivity predicted using the CLF model of this study compared to the model prediction of \citet{Bethermin:2010ea}.
The shaded region correspond the the $68\%$ confidence level from the MCMC model fits to the measurements used here.}\label{jnuz}
\end{figure}

\begin{figure}[htb!]
\epsfig{file=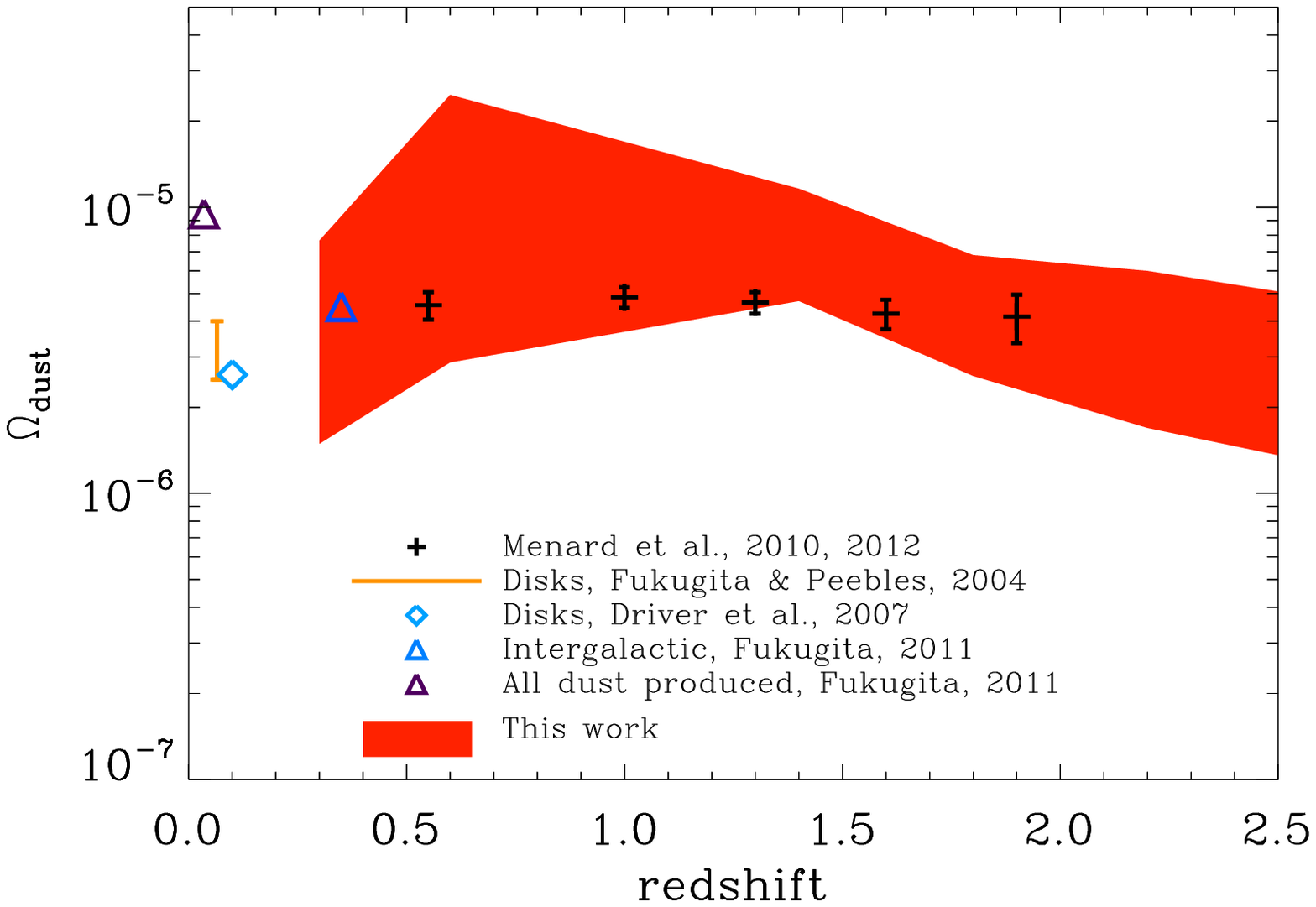,width=1\linewidth,clip=}
\caption{The cosmic density of dust $\Omega_{dust}$ vs redshift. The CLF model prediction (gray region) calculated with the best-fit luminosity functions
of Fig.~\ref{highz_plots_2} and compared to other works (see text).}\label{dust}
\end{figure}

We hence simplify the analysis as follows.  We first fit  the low-redshift parameters to the $z <0.2$ 250 $\mu$m
luminosity function measurements from \citet{Vaccari:2010vu} by only varying the $4$ parameters related to the
luminosity-halo mass relation and the duty cycle at low redshift, $\eta_{\rm DC}(z<0.2)$.  We assume no scatter in the luminosity-mass relation.
When fitting to high redshift data the total number of free parameters is
$11$: $3$ parameters for the $L-M$ relation ($\alpha_l^\prime$, $\beta_l^\prime$, $L_{0l}^\prime$, while $M_{0l,z<0.2}$ is kept fixed to the value found for the $z<0.2$ LF), plus the power index $p_M$ to account for the evolution of $M_{0l}$, $6$ duty cycle parameters and $1$ amplitude for the cirrus contamination, $C_{250}$.
%
%
When model fitting to the measured angular power spectrum at 250 $\mu$m we calculate the total theoretical $C_\ell$ as the sum of the $C_\ell$ for the
low-z HOD found in the previous fit and the $C_\ell$
calculated at $z>0.2$. This allows us to treat two types of galaxy populations that are contributing to the {\it Herschel} galaxy populatio, the low-$z$ ($z <0.1)$ dust in
late-type galaxies and the dusty spheroidal galaxies at high redshifts \citep{Lagache:2002xq}, and to account for a possible redshift evolution of the others L-M relation parameter 
($L_0$, $\alpha_l$, $\beta_l$). Here we consider galaxies brighter than $L>5\times10^7$ L$_{\odot}$ for the low redshift model, while 
we use $L>10^9$ L$_{\odot}$ to model-fit the high redshift data. These values are consistent with the flux cut of the galaxy 
samples considered. In order to account for the uncertainty in the exact value of $L_{\rm min}$ we 
have verified that an order of magnitude change in the value of $L_{\rm min}$ leads to  changes in the power spectra of the order of
5 to 6\% which is comparable to the $1\sigma$ error bars of the data. We find that the
 exact value of $L_{\rm min}$, within an order of magnitude, does not change the results considerably.

To model-fit the data we implement a Markov Chain Monte Carlo (MCMC) analysis using a modified version of the \texttt{cosmoMC} \citet{Lewis:2002ah} package.
The results for the low-z $z<0.2$ 250 $\mu$m luminosity function data are shown in Fig.~\ref{lowz_plots} and in Table~\ref{lowz_table}, where we show and tabulate the best-fit
to the luminosity function data and the best-fit values for the CLF parameters involved with the LF description, respectively.
In Fig.~\ref{lowz_plots} we show the HOD
calculated for the best-fit values of the parameters and its uncertainties.

The best-fit model to the angular power spectrum data and high-z luminosity functions is shown in Figures~\ref{highz_plots}-\ref{highz_plots_2}.
In Table \ref{highz_table} we tabulate the best-fit parameter values
(where the prime is used to distinguish the parameters for the high-z luminosity-mass relation from those for the low-z case) and in Fig.~\ref{contours}
we show the probability contours for the luminosity-mass relation parameters.

These results show that the model is able to fit the data even assuming no scatter in the luminosity mass relation. The HOD
shows a sharp cut-off at a mass of about $\log(M_{\rm min}/M_{\odot})\simeq10.8$ at $z=0$.  
This quantity could be compared to the threshold mass of the standard halo model and it is in agreement with the results of \citet{Amblard:2011gc}, where
it was found that $\log(M_{\rm min}/M_{\odot})\simeq11.5$ with a simple HOD for the dusty galaxies. However it should be 
noted that in this work we are not fitting directly the value of $M_{\rm min}$ as it is not a free parameter in our model.
Thus a direct comparison of our work to  \citet{Amblard:2011gc} may not be appropriate.

Both the values mentioned above are different from the recent results of \citet{Shang:2011mh}, where the authors used an improved version of the halo model
including a luminosity-mass relation to analyze the Planck-based
CFIRB anisotropy power spectrum \citep{Ade:2011ap}. They found that the most efficient halo mass scale for starformation is $M_{\rm eff}\simeq10^{12.65}M_{\odot}$,
which is closer to the typical value of optical galaxies in the standard halo model \citep{Cooray:2002dia, Abazajian:2004tn}.
Nevertheless, again, as already noted in \citet{Shang:2011mh}, the model used there is different from the one used in \citet{Amblard:2011gc} (and from the 
one used in this paper), and a direct comparison between  $M_{\rm eff}$ and  $M_{\rm min}$ may not be accurate. 
A proper comparison of our model to the results of \citet{Shang:2011mh} can be done through the
effective halo mass scale:
\begin{eqnarray}
M_{\rm eff}=\int dMn_h(M)M\frac{N_T}{ng}\ , .
\end{eqnarray}
\label{meff}
Integrating over our HODs,  we find $\log_{10}(M_{\rm eff})=12.63$ at $z=0$, which is comparable to the results of \citet{Shang:2011mh}. It is also comparable to 
 results for optical galaxies \citep{Cooray:2002dia, Abazajian:2004tn}. 
At $z=1.4$, where the emissivity peaks (see Fig. \ref{jnuz} and discussion below),  we find $M_{\rm eff}=11.45$. 

The contribution of satellite galaxies in our best-fit model becomes efficient at a mass scale $M_{\rm sat} \simeq17M_{\rm min}$ and the HOD of satellite galaxies
has a power-law behavior of $\propto M^s$ with the power-law slope of $s\simeq0.98$.
We show in Fig.~\ref{highz_plots} that this slope remains close to unity also for the high-z HOD where we find $s\simeq0.96$.
The relation $M_{\rm sat} \simeq17M_{\rm min}$  between minimum halo mass and the halo mass scale at which satellites appears and the slope of the satellite galaxies occupation number
 are consistent with expectations from numerical simulations and results obtained for optical galaxies \citep{Gao:2004au, Kravtsov:2003sg, Zheng:2004id, Hansen:2007fy, Shang:2011mh}.

Our result on the power-law slope with $s\sim0.98$ is
 different from \citet{Amblard:2011gc}  where a higher slope ($>1.6$) was found. The difference with $s \sim 1$ and $s > 1.6$ between the two works
comes in part from the fact that the model we have presented here accounts for that the brighter satellites are in more massive halos and in part 
from our rescaling of the cirrus amplitude. As discussed in \cite{Ade:2011ap}, \citet{Amblard:2011gc} overestimated the cirrus contamination and underestimates the
clustering power spectrum of the cosmic infrared background. We have refitted the cirrus amplitude as part of the joint fit to the power spectrum measurements using both a halo model
and a power-law power spectrum for cirrus.

The simple halo occupation number used in \citet{Amblard:2011gc} is not able to make a distinction between the luminosity and mass of satellite galaxies. 
Moreover, in the current description, the shape of the luminosity-mass relation determines the strength of the 1-halo term, while in the standard halo-model,
the strength of the 1-halo term is mainly determined by the slope of the halo occupation number of satellite galaxies.
As the 1-halo term is clearly detected in the {\it Herschel} CFIRB power spectrum the previous results were biased by an incorrect model that attempted
to model-fit high signal-to-noise power spectrum measurements. We are also finding an amplitude for the cirrus contamination which is smaller than $1$: $C_{250}=0.78\pm0.16$. A reduced cirrus contamination requires a reduced relative amplitude between the 1-halo and 2-halo terms that can be achieved with a lower slope for the satellite contributions. We note also that the fit to the cirrus amplitude confirms that the previous analysis of the same clustering data in \citet{Amblard:2011gc} overestimated the cirrus contribution as already has been found by the \citet{Ade:2011ap}, where a factor $\sim2$ of difference was found in the cirrus contamination. The model used in this work is hence able to alleviate the tension between Planck and Herschel analysis.

In terms of the duty cycle parameters, at $z < 0.2$,
where the constrain came only from the luminosity function data, we found a weak constraint on the duty cycle, that is $DC=0.54\pm0.48$ at $95\%$ confidence level. 
The amplitude of the luminosity function is in fact affected by both the duty cycle  and other parameters of the luminosity mass relation, in particular $L_0$ and $M_{0l}$.
Thus, in the absence of other constraints, the duty cycle can not be measured efficiently. On the other hand, 
the parameters of the $L-M$ relation also determine the HOD and the relative amplitude of 1-halo and 2-halo terms. Combining 
clustering measurements and luminosity function data can hence strongly improve the constraints on the duty cycle. 
Also the $z < 0.2$ luminosity function is better determined in narrow, deeper surveys compared to the case of the $z < 0.2$ LF.
All this results in a determination of high-z duty cycle parameters with  relatively small uncertainties.
We have found that the duty cycle generally decrease with redshift until $z\simeq1$ and slowly increase again for higher $z$ (see table \ref{highz_table} and Fig.~\ref{dc_z}).
The values are in the range $\eta_{\rm DC}\simeq0.16-0.4$ when $0.2<z<0.4$.

In \citet{Shang:2011mh} the authors have excluded very low duty cycle values ($\eta_{\rm DC}<0.05$) finding that the Planck power spectrum data favor
duty cycles close to unity. Our results lie in the middle between these two extreme cases.
However in \citet{Shang:2011mh} the duty cycle parameter is estimated by comparing the shot noise predicted for a fixed $\eta_{\rm DC}$
to the results of the empirical model of  \citet{Bethermin:2010ea}, rather than fitting to the data.
Moreover the model used here differs in the description of the luminosity mass-relation and introduces a possible redshift evolution both for the
duty cycle and for the 250 $\mu$m luminosity-halo mass relation. Interestingly our constraints  are much more similar to the results of \citet{Leeetal2009}, where the
model we are using here, with some differences in the L-M relation and its redshift dependence, was also
used to analyze the UV luminosity function and two point correlation function data of
starforming galaxies in the range $z=4-6$. In that work \citet{Leeetal2009} used $\eta_{\rm DC}$
as an input parameter rather than as a free parameters and
found that extremely short ($\eta_{\rm DC}<0.1$) and extremely long ($\eta_{\rm DC}>0.7$) duty cycles are ruled out at the $90\%$ confidence level. Our results also suggest a mid range for
$\eta_{\rm DC}$. The agreement could imply that a large fraction of the UV-selected starforming galaxy sample studied in \citet{Leeetal2009}  could
also be responsible for the CFIRB anisotropies. This could be directly tested via a cross-correlation between the two datasets and such studies are
expected in the near future given the {\it Herschel} imaging of some of the wide area legacy fields.

In Section~\ref{model} we described the duty cycle in terms of the
duration of the starformation $t_{\rm SF}$ in the halos with respect to the time interval $\Delta t$ covered by the survey. A long duty cycle $\eta_{\rm DC}\sim 1$ implies a
starformation time scale that is $t_{\rm SF}\gg\Delta t$, while small duty cycles with $\eta_{\rm DC}\sim 0$ correspond to the opposite case with $t_{\rm SF}\ll\Delta t$.
The central value with $\eta_{\rm DC}\sim0.5$ implies $t_{\rm SF}\simeq\Delta t$. In terms of the physical time, once accounted for the time interval spanned by each redshift bin,
the duty cycles listed in Table~\ref{highz_table} correspond to a starformation phase lasting for $t_{\rm SF}\simeq0.3-1.6$ Gyr.
Such a long starformtaion timescale is consistent with what has been suggested in \citet{Lapi:2011ca} and the physical model of \citet{Granato:1999hn, Granato:2003ch}.
Such a long time scale rule out models where the CFIRB is dominated by gas-rich mergers with $t_{\rm SF}\simeq10-100$ Myr.

It is worth noticing that in this analysis we are assuming a duty cycle that is independent of mass. It could very well be that
the duration of starformation depends on the halo mass.
Given the large number of free parameters in the analysis we are not able to parametrize a possible mass or luminosity dependence of $\eta_{\rm DC}$,
but we regard this possibility as a future improvement to this model.

Our analysis suggests that the L-M relation has a redshift dependence, with $M_{0l}$ that decreases for higher redshifts (see eq.~\ref{mol_evolution}).
Decreasing $M_{0l}$ is equivalent to increase the characteristic luminosity of the luminosity function.
The fit to the high-z luminosity function data is shown in Fig.~\ref{highz_plots_2}.
In particular we find $M_{0l}\propto (1+z)^{-4.32\pm0.09}$.  Although a direct comparison is complicated because of the very different models used, we note
that this result is similar to the evolution seen in \citet{LeFloc'h:2005pd}, where the characteristic luminosity has been found to have a redshift dependence $\propto(1+z)^{3.2^{+0.7}_{-0.2}}$.

The total luminosity-mass relation calculated as:
\begin{eqnarray}
L(M)=\bar{L}(M)+\int N(m|M)\bar{L}(m)dM ,
\end{eqnarray}
is shown in Figure \ref{LM}. The shaded regions represent the $1\sigma$ uncertainty and it can be seen that the data are able to constrain the luminosity-mass relation with good precision even
for the $z<0.2$ case using only the low redshift luminosity function data. 
The luminosity-mass relations we are finding show a linear behavior. However the luminosity functions are steep at the low-faint luminosity end.
 This result in a tension when the observed turnover is attempted to be explained through 
the abundance matching approach in \citet{Bethermin2012}. We consider this a natural 
consequence of the fact that in this work we are attempting to fit simultaneously datasets in a large range of redshifts together with anisotropy power spectrummeasurements. 
Moreover we observe that there is no clear visible turnover in the data that we are fitting 
without imposing any prior or constraint on the faint-end of the luminosity functions. The faint-end description, both in data and in models, should be
further improved. 

In Figure \ref{jnuz} we show the emissivity corresponding to the best-fit model, calculated according to equation (\ref{jnu}) and compared to the emissivity of the parametric model of \citet{Bethermin:2010ea} (see also \citealt{Penin2012}). The extended tail at $z>3$ is due to
the constant and the fast increase of the luminosity function with redshift.
We have verified that using the emissivity of \citet{Bethermin:2010ea} implies a few percent difference in the best-fit values of the CLF parameters in the
model presented here, comparable to the $1\sigma$ error bars. Future analysis may require however a different redshift parameterization of the average luminosity-mass relation.

We show also the satellite fraction for the four high redshift bins calculated as \citet{vandenBosch:2006iq}:
\begin{eqnarray}
f_{sat}(L)=\frac{1}{\phi(L)}\int_{M^{\prime}}^{\infty} dMP(L|M)n_T(M)\, ,
\end{eqnarray}
where $M^{\prime}$ is the mass scale where there is one galaxy brighter than $L$. The satellite fraction is an important test for galaxy formation models and to establish the properties of galaxy-halo relation. We find that the fraction is about $22-25\%$ at $L=10^9 L_{\odot}$ and decreases quickly to less than $5\%$ at $10^{11} L_{\odot}$ while we don't find a significant redshift dependence.
The decreasing behavior with mass is due to the fact that satellite galaxies at a given luminosity are located in more massive (and hence less numerous) halos with respect to central galaxies. This result for $f_{sat}$ is also in agreement with \citet{vandenBosch:2006iq, cooray2006,Coupon:2011mz}.

Finally in Fig.~\ref{dust} we show the fraction of dust with respect to the critical density of the Universe $\rho$, calculated as
	\begin{eqnarray}\label{omegadust}
\Omega_{\rm dust}=\frac{1}{\rho_0}\int_{L_{min}} dL\phi(L,z)M_{dust}(L) \, ,
\end{eqnarray}
where $M_{dust}$ is the dust mass corresponding to a given IR luminosity and we use Eq.~4 in \citet{Fu:2012ba}. The results are compared to those in Fig.~7 of \citet{Menard:2012am} where $\Omega_{\rm dust}$ has been determined with reddening of metal-line
absorbers. In Fig.~\ref{dust}  we also show other estimates of the mass density of dust as summarized by \citet{Menard:2012am} from \citet{Fukugita2004, Driver2007, Menard2010, Fukugita2011}. 
We have combined the points from \citet{Menard2010} for the dust contributions of halos and those from \citet{Menard:2012am} in a single set of data points, under the assumption that the 
amount of dust in halos doesn't evolve significantly with redshift.
 We parameterize the opacity with a power low $k_d\propto \nu^{\beta_d}$ with the power index in the range $\beta_d=1.5-2$. The calculation requires the spectral energy distribution of dust and we assume a thermal black-body spectrum wit
dust temperature in the range $T=25-35K$. We allow for a large range in dust temperature, taken as a uniform prior, to allow for the
range of  values seen in current data \citep{Amblard2010}. In equation (\ref{omegadust}) we integrate over luminosities $L_{min}>10^9L_{\odot}$. However in this calculation the choice of $L_{min}$ is less relevant, since the uncertainty on $\Omega_{\rm dust}$ is dominated by the large range of temperatures and spectral indices considered.
The gray region correspond to the prediction for these parameters ranges using the best-fit luminosity functions of Fig.~\ref{highz_plots_2}.

%

\section{Conlcusion}

We have presented an analysis of the {\it Herschel}-SPIRE CFIRB power spectrum at 250 $\mu$m and
the luminosity functions up to $z=4$.  We use a conditional luminosity function approach to model the far-IR bright galaxies.
We have modeled the 250 $\mu$m luminosity function and its evolution with redshift introducing a redshift dependent duty-cycle parameter.
This description represents an improved version of the halo-model that offers an advantage by accounting for the luminosity dependence of the
satellite galaxies as a function of the halo mass. The underlying ingredient is the
 galaxy luminosity-halo mass relation.

We have found that current {\it Herschel} data are able to constrain the model despite the high number of free parameters.
The results of our analysis indicate that the cosmic far-IR background is dominated by starforming galaxies in an extended phase
of starformation rather than bright starbursts that are fueled by gas-rich mergers.
 We found duty cycles corresponding to a dusty starformation phase
lasting $\sim0.3-1.6$ Gyr, which is in agreement with previous analysis of
starforming UV-selected galaxies at high redshifts.

We have also found  that the halo occupation number for satellite galaxies has a power-law slope that is about 0.98
over the redshift range $0 < z < 4$.
This solves the tension between previous analysis of the same {\it Herschel} power spectrum data and other determinations of the halo occupation number for
galaxies in the literature.
Finally we have estimated the cosmic density of dust residing in the dusty, starforming galaxies responsible for the cosmic far-IR background anisotropies to
be $\Omega_{\rm dust} \sim 3\times10^{-6}$ to $2 \times10^{-5}$.

\section{Acknowledgments} 	
We thank Alexandre Amblard for useful communications and Brice Menard for clarifying the SDSS dust measurements.
We acknowledge support from NSF CAREER AST-0645427, Herschel funding from NASA Herschel Science Center through a contract from JPL/Caltech,
and NASA ADAP award NNX10AD42.
\newpage
\vspace{50pt}

\end{document}